\documentclass[aps,twocolumn]{revtex4}
\usepackage[utf8]{inputenc}
 \usepackage{amsmath,amssymb}
\usepackage{color}
\usepackage{graphicx}
\usepackage{epstopdf}                             
\usepackage{float}
 \usepackage{lineno}
\newcommand{\be}{\begin{equation}}
\newcommand{\ee}{\end{equation}}

\begin{document}

\title{Slow-roll inflation in (dual) Kaniadakis cosmology}
\author{Leila Liravi and Ahmad Sheykhi \footnote{asheykhi@shirazu.ac.ir}}
\address{Department of Physics, College of
Science, Shiraz University, Shiraz 71454, Iran\\
Biruni Observatory, College of Science, Shiraz University, Shiraz
71454, Iran}

\begin{abstract}
We investigate slow-roll inflation within the framework of
Kaniadakis and dual Kaniadakis cosmology, where the usual entropy
formalism is generalized through a deformation parameter $\kappa$.
By deriving the modified Friedmann equations and the corresponding
inflationary dynamics induced by Kaniadakis entropy, we analyze
the deviations from standard inflation arising from
$\kappa$-corrections.To test the viability of these models, we
evaluate two representative inflationary potentials: the power-law
and the Mexican-hat (symmetry breaking) potentials. We compute the
scalar and tensor spectral indices, the tensor-to-scalar ratio,
and the primordial power spectrum, comparing them against the
latest Planck and BICEP observational constraints. Our results
show that in the standard Kaniadakis formulation, viable slow-roll
inflationary scenarios compatible with observations can be
obtained, although the allowed values of the deformation parameter
are  suppressed, yielding bounds of $\kappa \sim
\mathcal{O}(10^{-9})$ and $\kappa \sim \mathcal{O}(10^{-37})$ for
the power-law and Mexican-hat potentials, respectively.
Conversely, for the dual Kaniadakis formulation, we find no
physical parameter space that simultaneously satisfies the
observational bounds for these potentials, rendering this
formulation phenomenologically disfavored. These findings suggest
that while standard Kaniadakis cosmology can leave potentially
observable imprints on primordial perturbations, the constraints
on $\kappa$ during inflation are significantly weaker than those
from late-time cosmological bounds, hinting at a possible
energy-scale dependence of the non-extensive deformation
parameter.
\end{abstract}
 \maketitle
 \newpage
\section{Introduction\label{Intro}}
Inflationary cosmology provides a compelling paradigm for
understanding the early stages of the universe, resolving
long-standing puzzles such as the horizon and flatness problems
while also seeding the primordial density perturbations that later
grew into large-scale structures. The simplest models of inflation
rely on a scalar field slowly-rolling down a potential, leading to
an almost exponential expansion of space. While successful, the
standard inflationary framework is based on classical and
semiclassical assumptions tied to Boltzmann-Gibbs (BG)
thermodynamics.

In recent years, alternative approaches have gained traction,
particularly those informed by non-extensive statistical
mechanics, which introduces novel entropy measures that can modify
the dynamics of cosmological models. Among the generalizations of
BG entropy, the so called Kaniadakis entropy provides a
one-parameter deformation of the standard entropy, controlled by a
dimensionless parameter $\kappa$, which accounts for relativistic
and non-extensive effects in high-energy regimes \cite{Kan1,Kan2}.
When $\kappa\rightarrow 0$, the Kaniadakis entropy reduces to the
BG entropy, but introduces power-law corrections for
$\kappa\neq0$. This offers a  new window into quantum
gravitational or high-energy thermodynamic effects. Recently, this
formalism has been extended to cosmology, where the Kaniadakis
entropy modifies the horizon thermodynamics and consequently the
dynamical equations governing the evolutions of the universe. A
related but distinct framework is dual Kaniadakis cosmology, where
the deformation parameter takes an imaginary value
$\kappa\rightarrow i \kappa$, leading to oscillatory corrections
that may have distinct phenomenological implications.

In this work, we explore slow-roll inflation within the context of
(dual) Kaniadakis cosmology, where the modified entropy-area
relation alters the Friedmann equations and the dynamics of
inflation. We derive the slow-roll parameters and analyze their
dependence on the $\kappa$-deformation, examining whether
Kaniadakis corrections can leave observable imprints on the
inflationary power spectrum or relax fine-tuning constraints. Our
results aim to bridge the gap between non-extensive thermodynamics
and primordial cosmology, offering insights into how
quantum-gravitational statistics might reshape the inflationary
paradigm. Our work differs from \cite{Lamb} in that, we consider
not only Kaniadakis cosmology, but also the case of \textit{dual}
Kaniadakis cosmology which bring rich physics. In particular,
using the entropic force scenario proposed by Verlinde \cite{Ver},
and in the non-relativistic Newtonian gravity, it has been shown
that the \textit{dual} Kaniadakis entropy can provide a
theoretical origin for the Modified Newtonian Dynamics (MOND)
theory \cite{Ava2}. This implies that one is capable to address
the flat-rotation curves of spiral galaxies through
thermodynamics-gravity conjecture \cite{Ava2}. This may also
change the profile of the slow-roll inflation and affects the
scalar and tensor spectral indices, the tensor-to-scalar ratio.
Besides, in \cite{Lamb}, the authors consider a simple linear
potential $V(\phi)\sim \phi$, while here we extend the studies to
more general potentials such as $V(\phi)\sim \phi^n$ and Mexican
hat potential, $V(\phi)=\lambda_{0}( \phi^{2}-\phi_{0}^{2})^2$.
This would much constrain the model parameter $\kappa$ from
inflationary arguments. Given these modifications, it is natural
to explore how slow-roll inflation is altered in (dual) Kaniadakis
cosmology. Previous studies on Kaniadakis modified cosmology have
examined the effects of parameter $\kappa$ on dynamics of the
universe \cite{Lym,Her,ShKa1,ShKa2}, dark energy models
\cite{Dre,Kumar,Almada}, primordial Big-Bang Nucleosynthesis (BBN)
\cite{Luciv,Ava2}. However, the implications of the Kaniadakis
modified cosmology for the early cosmic inflation-particularly the
spectral tilt $n_s$ and $r_s$, tensor-to-scalar ratio, and running
of the spectral index-remain less explored \cite{Lamb}. Since
inflationary observables are tightly constrained by Cosmic
Microwave Background (CMB) data, any deviation from standard
predictions could either constrain the $\kappa$ parameter or
provide a signature of non-extensive thermodynamics in the early
universe.

The organization of our paper is as follows. In the next section,
we review (dual) Kaniadakis entropy and derive the modified
Friedmann equations.  In Section III, we analyze slow-roll
inflation in the context of modified (dual) Kaniadakis cosmology,
computing the inflationary observables. In section IV, we study
power spectrum and phenomenological consistency of our model.
Finally, in section V, we summarize our conclusions and outline
future directions.
\section{Modified Friedmann equations through (dual) Kaniadakis entropy \label{Kan}}
In this section, we review the origin and formalism of the
generalized Kaniadakis entropy and present the modified
cosmological Friedmann equations based on the Kaniadakis entropy.
Kaniadakis entropy is one-parameter entropy which generalizes the
classical Boltzmann-Gibbs-Shannon entropy. It originates from a
coherent and self-consistent relativistic statistical theory. The
advantages of Kaniadakis entropy are that it preserves the basic
features of standard statistical theory, and in the limiting case
restore it \cite{Kan1,Kan2}. The general expression of the
Kaniadakis entropy is given by \cite{Kan1,Kan2}
\begin{eqnarray}
S_{\kappa}=- k_{_B} \sum_i n_i\, \ln_{_{\{{\scriptstyle \kappa}\}}}\!n_i  ,
\end{eqnarray}
with $k_{_B}$ is the Boltzmann constant, and
\begin{eqnarray}
\ln_{_{\{{\scriptstyle \kappa}\}}}\!x=\frac{x^{\kappa}-x^{-\kappa}}{2\kappa}.
\end{eqnarray}
Here $\kappa$ is called the Kaniadakis parameter which is a
dimensionless parameter ranges as $-1<\kappa<1$, and measures the
deviation from standard statistical mechanics. In the limiting
case where $\kappa\rightarrow0$, the standard entropy is restored.

In such a generalized statistical theory the distribution function
becomes \cite{Kan1,Kan2}
\begin{eqnarray}
n_i= \alpha \exp_{_{\{{\scriptstyle \kappa}\}}}[-\beta (E_i-\mu)] ,
\end{eqnarray}
 where
\begin{eqnarray}
&& \exp_{_{\{{\scriptstyle \kappa}\}}}(x)=
\left(\sqrt{1+\kappa^2x^2}+\kappa x\right)^{1/\kappa}, \\
&&\alpha=[(1-\kappa)/(1+\kappa)]^{1/2\kappa},\\
&&\beta^{-1}=k_{_{B}}T \sqrt{1-\kappa^2}.
\end{eqnarray}
Let us note that the chemical potential $\mu$ can be fixed by
normalization \cite{Kan1,Kan2}. Alternatively, Kaniadakis entropy
can be expressed as
\cite{Abreu:2016avj,Abreu:2017fhw,Abreu:2017hiy,Abreu:2018mti,Yang:2020ria,
Abreu:2021avp}
\begin{equation}
 \label{kstat}
S_{\kappa} =-k_{_B}\sum^{W}_{i=1}\frac{P^{1+\kappa}_{i}-P^{1-\kappa}_{i}}{2\kappa}.
\end{equation}
Here  $P_i$ is the probability in which the system to be in a
specific microstate and $W$ represents the total number of the
system configurations. Throughout this paper we set
$k_{_B}=c=\hbar=1$.

It is also interesting to apply the  Kaniadakis entropy to the
black hole thermodynamics. It is well known that the entropy of
the black hole, in Einstein gravity, obeys the so called
Bekenstein-Hawking entropy, which states that the entropy of the
black hole horizon is proportional to the area of the horizon,
$S_{BH}= A/(4G)$. Now we assume  $P_i=1/W$, and using the fact
that BG entropy is $S\propto\ln(W)$, and $S=S_{BH}$, we get
$W=P^{-1}_{i}=\exp\left[S_{BH}\right]$ \cite{Mor}.

Substituting $P_i=e^{-S_{BH}}$ into Eq. (\ref{kstat}) we arrive at
 \begin{equation} \label{kentropy}
S_{\kappa} = \frac{1}{\kappa}\sinh{(\kappa S_{BH})}.
\end{equation}
When $\kappa\rightarrow 0$ one recovers the standard Bekenstein-Hawking
entropy, $S_{\kappa\rightarrow 0}=S_{BH}$. Considering the fact that
deviation from the standard Bekenstein-Hawking is small, we expect
that $\kappa\ll1$. Therefore, we can expand expression (\ref{kentropy})
as
\begin{equation}\label{kentropy2}
S_{\kappa}=S_{BH}+ \frac{\kappa^2}{6} S_{BH}^3+ {\cal{O}}(\kappa^4).
\end{equation}
The first term is the usual area law of black hole entropy, while
the second term is the leading order Kaniadakis correction term.
In what follows, we shall apply the above expression to extract
the modified friedmann equations.
An alternative generalization, known as the \textit{dual} or
\textit{deformed} Kaniadakis entropy, can be introduced by
inverting the relation between the black hole microstate number
$W$ and entropy. Starting from the expression
\begin{equation} \label{SKD}
S_{BH} = \frac{W^\kappa - W^{-\kappa}}{2\kappa},
\end{equation}
one can solve for $W$ to obtain
\begin{equation} \label{Wdual}
W = \left(\kappa S_{BH} \pm \sqrt{1 + \kappa^2
S_{BH}^2}\right)^{1/\kappa},
\end{equation}
where we choose the positive root to ensures the correct
Boltzmann-Gibbs limit as $\kappa \to 0$. The dual Kaniadakis
entropy is then defined as \cite{AbreuNet1,Amb}
\begin{equation} \label{SKD1}
S^*_{\kappa} = \ln W = \frac{1}{\kappa} \ln \left(\kappa S_{BH} +
\sqrt{1 + \kappa^2 S_{BH}^2}\right),
\end{equation}
which reduces to the standard Bekenstein-Hawking entropy in the
limit $\kappa \rightarrow 0$. For small $\kappa \ll 1$, a series
expansion gives
\begin{equation} \label{SKD2}
S^*_{\kappa} = S_{BH} - \frac{\kappa^2}{6} S_{BH}^3 +
\mathcal{O}(\kappa^4),
\end{equation}
where the first term is the usual area law and the second term is
the leading-order correction. Unlike the standard Kaniadakis
entropy, the dual entropy has a negative sign in the leading
correction. Thus, both Kaniadakis and dual Kaniadakis entropies
can be written, up to first-order corrections, as
\begin{equation} \label{SKDG}
S_{\kappa} = S_{BH} \pm \frac{\kappa^2}{6} S_{BH}^3 +
\mathcal{O}(\kappa^4),
\end{equation}
with $+$ for standard Kaniadakis and $-$ for the dual formalism.
\subsection{Corrections to Friedmann equations}
Kaniadakis entropy, introduces corrections to the Friedmann
equations, adding terms that can be interpreted as an effective
dark energy component. Modified cosmology through Kaniadakis
entropy~\eqref{kentropy2} has been explored in~\cite{Lym}, where
it was shown that such corrections can lead to new cosmological
behaviors and deviations from the standard $\Lambda$CDM scenario.
A geometric interpretation of these corrections has also been
provided in~\cite{ShKa1,ShKa2}, highlighting that entropy
modifications should correspondingly modify the geometry of the
field equations.

Similarly, the dual (deformed) Kaniadakis entropy provides an
alternative formulation, with the leading-order correction
appearing with an opposite sign. Considering both the standard and
dual forms allows us to explore a broader range of cosmological
scenarios and study how these entropy corrections influence the
dynamics and evolution of the Universe in a clear and systematic
way. Our aim here is to review derivation of the modified
Friedmann equations through Kaniadakis entropy. We follow the
approach presented in \cite{Kan1}.

Our starting point is a homogeneous and isotropic flat FRW metric
\be \label{FRW} ds^2 = h_{\mu\nu} dx^\mu dx^\nu + \tilde r^2
(d\theta^2 + \sin^2\theta\, d\phi^2), \ee where $\tilde r =
a(t)\,r$, $x^0 = t$, $x^1 = r$, $h_{\mu\nu} = \mathrm{diag}(-1,
a^2)$, and $a(t)$ is the scale factor. By treating the Universe as
a spherical thermodynamic system, the radius of the apparent
horizon can be expressed as $\tilde r_A = 1/\sqrt{H^2+k/a^2}$,
where $H$ denotes the Hubble parameter. The corresponding
temperature is determined through the surface
gravity~\cite{Akbar:2006kj}, and is given by \be \label{temp} T_h
= -\frac{1}{2\pi \tilde r_A} \left(1 - \frac{\dot{\tilde r}_A}{2 H
\tilde r_A}\right). \ee Assuming a perfect fluid with energy
density $\rho$ and pressure $p$, the energy-momentum tensor reads
\be T_{\mu\nu} = (\rho + p) u_\mu u_\nu + p g_{\mu\nu},
 \ee
and the continuity equation becomes \be \label{ce} \dot \rho = -3
H(\rho + p). \ee Using the gravity-thermodynamic conjecture, the
first law of thermodynamics on the apparent horizon is \be
\label{flt} dE = T_h dS_h + W dV, \ee where $E = \rho V$ is the
total energy enclosed by the apparent horizon, with
$V=\frac{4\pi}{3}\tilde r_A^{3}$ is the volume enveloped by a
3-dimensional sphere with the area of apparent horizon $A=4\pi
\tilde r_A^{2}$.

The work density is defined as \cite{Hay2} \be \label{work} W =
-\frac{1}{2} T^{\mu\nu} h_{\mu\nu} = \frac{1}{2} (\rho - p). \ee
Omitting the detailed algebraic steps-although one can follow the
details of calculation in \cite{ShKa1}- we assume that the entropy
associated with the apparent horizon follows the Kaniadakis
form~\eqref{kentropy2}. Substituting relations \eqref{kentropy2},
\eqref{temp} and \eqref{work} into the first law of
thermodynamics~\eqref{flt}, one arrives, in a flat universe, at
the corresponding modified Friedmann equations~\cite{ShKa1}
\begin{eqnarray}\label{FFE}
 &&H^2 - \kappa^2 \frac{\pi^2}{2 (G H)^2} =
\frac{8\pi G}{3} \rho,\\
&&\dot H \left[ 1 + \kappa^2 \frac{\pi^2}{2 (G H^2)^2} \right] =
-4 \pi G (\rho + p). \label{SFE}
\end{eqnarray}
Clearly these equations reduce to the standard Friedmann equations
when $\kappa \to 0$. These represent the leading-order corrections
in Kaniadakis cosmology inspired by the modified entropy
\eqref{kentropy2}.

Similarly, one can consider the dual (deformed) Kaniadakis
entropy~\eqref{SKD2} as the apparent horizon entropy. Following
the same thermodynamic procedure, the dual entropy introduces
correction terms of opposite sign, leading to \cite{Ava2}
\begin{eqnarray} \label{FFEdual}
&&H^2 + \kappa^2 \frac{\pi^2}{2 (G H)^2} = \frac{8\pi G}{3}
\rho,\\
&&\dot H \left[ 1 - \kappa^2 \frac{\pi^2}{2 (G H^2)^2} \right] =
-4 \pi G (\rho + p).\label{SFEdual}
\end{eqnarray}
Thus, the dual Kaniadakis entropy naturally provides an
alternative cosmological scenario, with leading-order corrections
of reversed sign compared to the standard case.
\section{Slow-roll inflation in (dual) Kaniadakis cosmology \label{Inflation}}
We investigate slow-roll inflation in the context of (dual)
Kaniadakis-modified Friedmann cosmology. Following the approach
in~\cite{Keskin,LucBarInf}, we focus on the early high-energy
phase of the Universe under the slow-roll regime~\cite{OdSR},
assuming that its evolution is driven by a scalar field $\phi$
with potential $V(\phi)$. The key observables characterizing
inflation in this framework are the scalar spectral index $n_s$
and the tensor-to-scalar ratio $r$, defined for a minimally
coupled scalar field as in~\cite{Inf1,Inf2}. In what follows, both
the scalar spectral index $n_s$ and the tensor-to-scalar ratio $r$
will be analyzed in detail within the considered framework. For
completeness, we briefly review the standard slow-roll results
before introducing the corresponding modifications arising from
the (dual) Kaniadakis framework.

In the following analysis, we consider two inflationary potentials
and treat each of them independently. For every potential, we
first present the standard slow-roll formulation and then derive
its Kaniadakis-modified (dual) version.
\subsection{The case with potential $V(\phi)=V_{0}\phi^{n}$}
\subsubsection{ Standard slow-roll inflation  }
Within the standard slow-roll framework, we consider a canonical
scalar field $\phi$ with potential $V(\phi)$ in a flat
Friedmann-Lema\^{i}tre-Robertson-Walker (FLRW) universe. The
energy density and pressure associated with the inflation field are
expressed as
\begin{eqnarray}
    \label{rho}
    &&\rho_\phi = \frac{1}{2}\dot{\phi}^2 + V(\phi),\\
    &&p_\phi = \frac{1}{2}\dot{\phi}^2 - V(\phi).  \label{p}
\end{eqnarray}
The background dynamics are governed by the Friedmann equations,
which, for a homogeneous field, reduce to
\begin{equation} \label{e9}
H^2= \frac{8\pi G}{3} \left( \frac{1}{2}\dot{\phi}^2 + V(\phi)
\right),
\end{equation}
\begin{equation}\label{e10}
\dot{H} = -4\pi G \dot{\phi}^2.
\end{equation}
The scalar field obeys the Klein-Gordon equation,
\begin{equation}
\label{KG}
\ddot \phi + 3H \dot\phi + \partial_\phi V(\phi) =0\,,
\end{equation}
where $\partial_\phi V\equiv\frac{dV}{d\phi}$. Under the
assumption that the potential energy of the scalar field dominates
over its kinetic energy, the slow-roll conditions are
~\cite{Inf1,duality}
\begin{equation}
\label{con} \ddot \phi \ll H \dot \phi, \qquad
\frac{1}{2}\dot\phi^2 \ll V(\phi),
\end{equation}
ensuring that the field rolls slowly down its potential.

The slow-roll parameters are defined as \cite{Lamb}
\begin{eqnarray}
    \label{slow1}
    \epsilon&=&- \frac{\dot H}{H^2}\,,\\[2mm]
    \eta&=&-\frac{\ddot H}{2H\dot H}\,.
    \label{slow2}
\end{eqnarray}
and the main inflationary observables, the scalar spectral index
$n_s$ and the tensor-to-scalar ratio $r$, are given by
\cite{Baumann}
\begin{eqnarray}
    \label{r}
    r&=&16\epsilon\,,\\[2mm]
    n_s&=& 1-4\epsilon+2\eta\,.
    \label{ns}
\end{eqnarray}
If we apply condition (\ref{con}) into Eq. (\ref{e9}), we find
\begin{align}
    \label{fr2}
    H &\simeq
  \sqrt{\frac{8\pi G }{3}V(\phi)} \,.
\end{align}
Using Eqs. (\ref{e10}) and (\ref{fr2}), the parameters
(\ref{slow1}) and (\ref{slow2}), take the form
\begin{align}\label{eps22}
\epsilon \simeq \frac{3}{2V} \, \dot{\phi}^2 \,,
\end{align}
\begin{align}
        \label{eps223}
    \eta \simeq - \frac{1}{\dot{\phi}} \sqrt{\frac{3}{8\pi G V}} \, \ddot{\phi} \,.
\end{align}
As discussed in~\cite{Keskin}, both slow-roll parameters should be
evaluated at the time of horizon crossing, when the inflaton
fluctuations stop evolving and become constant.

The overall duration of the inflationary expansion is
characterized by the number of e-folds, $N$~\cite{Carrol}, which
is defined as
\begin{align}
\label{N}
N = \int_{t_i}^{t_f} H(t)\, dt\,,
\end{align}
where $t_i$ and $t_f$ represent the beginning and the end of the
inflationary phase, respectively. Since the observable properties
of primordial perturbations are determined at the horizon
crossing, it is convenient to set the initial time of inflation to
this moment, i.e. $t_i = t_c$. In terms of the inflaton field,
this corresponds to $\phi_i = \phi_c$.
\\With this identification, Eq.~\eqref{N} can be rewritten as
\begin{align}
\label{efold1}
N = \int_{\phi_c}^{\phi_f} \frac{H}{\dot{\phi}}\, d\phi\,,
\end{align}
where $\phi_c = \phi(t_c)$ is the value of the inflaton field at
horizon crossing, and $\phi_f = \phi(t_f)$ denotes its value at
the end of inflation.

Now we focus on a power-law potential $V(\phi)=V_0 \phi^n$ for the
inflaton field as a concrete example to illustrate the
inflationary dynamics. In this work, we adopt a
potential-motivated approach, meaning that the potential is
specified first and then the corresponding Friedmann equations are
solved. Indeed, the potential $V$ affects the dynamics of the
model through the scalar field energy density and pressure, as
defined in Eqs.~\eqref{rho} and~\eqref{p}, respectively. These
quantities contribute to the total energy density and pressure
appearing on the right-hand side of the modified Friedmann
equations.
 Under the slow-roll approximation, the scalar field dynamics simplifies to
\begin{align}
\label{phidot1}
\dot{\phi} \simeq - \frac{1}{3H}\,\partial_\phi V.
\end{align}
This allows for an analytical evaluation of inflationary observables
such as \(n_s\) and \(r\) as well as the number of e-folds, without requiring the exact values of
the potential parameters. By using the definition of the potential and Eq.~\eqref{fr2}, we obtain
\begin{align}
\label{phidot 2} \dot\phi\simeq-\frac{n}{2}\sqrt{\frac{V_0}{6\pi
G}}\phi^{(n-2)/2}.
\end{align}
We define the end of inflation (and the corresponding number of
e-folds $N$) as the moment when $\epsilon(\phi_f) \sim 1$. To
proceed, we first rewrite Eq.~\eqref{eps22} as
\begin{align}
\label{eps2}
 \epsilon(\phi_f)\simeq\frac{n^2}{16\pi
 \hspace{0.2mm}G\hspace{0.2mm}\phi_f^2}.
\end{align}
Solving with respect to $\phi_f$, we obtain
\begin{align}
\label{phif12} \phi_f\simeq\frac{n}{4\sqrt{\pi G}}.
\end{align}
The inflaton value at horizon crossing, $\phi_c$, can subsequently
be obtained from Eq.~\eqref{efold1}, yielding
\begin{align}
\label{ansatz}
N\simeq -\frac{n}{4}+\frac{4\pi \hspace{0.2mm}G \phi_c^2}{n}\,.
\end{align}
Solving the above equation for $\phi_c$, we find
\begin{align}
\label{phic}
\phi_c\simeq\frac{1}{4}\sqrt{\frac{n(4N+n)}{\pi\hspace{0.2mm}G}}\,.
\end{align}
This relation allows to rewrite the tensor-to-scalar
ratio~\eqref{r} and the scalar spectral index~\eqref{ns} as
functions of the number of e-folding, $N$. To this end, we note
that
\begin{eqnarray}
\epsilon(\phi_c)&\simeq&\frac{n}{4N+n},
\end{eqnarray}
\begin{eqnarray}
\label{etas} \eta(\phi_c)&\simeq&\frac{n-2}{4N+n},
\end{eqnarray}
which can then be substituted into Eqs.~\eqref{r} and~\eqref{ns}
to yield
\begin{equation}
r=\frac{16n}{4N+n}\,.
\end{equation}
\begin{equation}
n_s = 1-\frac{2(n+2)}{4N+n}.
\end{equation}
\subsubsection{ Kaniadakis-modified slow-roll inflation}
We now examine inflation within the (dual) Kaniadakis-modified
Friedmann framework. If we solve the first modified Friedmann
equations~\eqref{FFE} and \eqref{FFEdual} with respect to $H$ and
use the second conditions in \eqref{con}, we obtain to the leading
order in $\kappa$
\begin{eqnarray}
\label{Happ} H\simeq\sqrt{{\frac{8\pi G}{3}\hspace{0.2mm}V}}\pm
\sqrt{\frac{27\pi}{2\hspace{0.2mm}G^{7}\hspace{0.2mm}V^{3}}}\frac{\kappa^2}{64}\,,
\end{eqnarray}
where the plus and minus signs correspond to the Kaniadakis and
dual formulations, respectively. In addition, using the second
Friedmann equations~\eqref{SFE} and ~\eqref{SFEdual} together with
Eq. \eqref{rho} and  Eq. \eqref{p}, we obtain
    \begin{eqnarray}
    \label{hdot3}
    \dot H\simeq\left(-4\pi G\pm\frac{9\pi\kappa^2}{32G^3V^2}\right)\dot\phi^2\,.
\end{eqnarray}
As a result, the parameters in Eqs. \eqref{slow1} and
\eqref{slow2} take the following form
\begin{eqnarray}
    \label{eps}
    \hspace{-2mm}\epsilon\hspace{-0.5mm}\simeq\hspace{-0.5mm}\left(\frac{3}{2V}\mp\frac{27\kappa^2}{128}\frac{1}{G^4V^3}\right)\dot\phi^2\,,
\end{eqnarray}
\begin{eqnarray}    \label{eta}
    \hspace{-2mm}\eta\hspace{-0.5mm}\simeq\hspace{-0.5mm}\left[-\sqrt{\frac{3}{8\pi G\hspace{0.2mm}V}}\,\ddot\phi\right.
    \pm\left.\sqrt\frac{3}{2\pi\hspace{0.2mm}G^9\,V^7}\frac{9\kappa^2}{512}\left(\ddot\phi\,V-2\dot\phi^2\,\partial_\phi
    V\right)\right]\frac{1}{\dot\phi},\nonumber\\
\end{eqnarray}
where hereafter the upper and lower signs in the equations
correspond to the Kaniadakis and dual formulations, respectively.
If we substitute the definition of the scalar field potential and
Eq.~\eqref{Happ} in Eq.~\eqref{phidot1}, we obtain
\begin{eqnarray}
        \label{phii}
    \dot\phi\simeq-\frac{n}{2}\sqrt{\frac{V_0}{6\pi G}}\phi^{(n-2)/2}
\pm\frac{3\hspace{0.2mm}n\kappa^2}{512}\sqrt{\frac{3}{2\pi\left(G^9\hspace{0.3mm}V_0^3\right)}}\phi^{-(2+3n)/2}.\nonumber\\
\end{eqnarray}
Note that inflation (and e-fold counting $N$) ends when
$\epsilon(\phi_f) \simeq 1$. Using the above equation,
Eq.~\eqref{eps} can be rewritten as
\begin{eqnarray}\label{epsphi1}
\epsilon(\phi_f)\simeq\frac{n^2}{16\pi
\hspace{0.2mm}G\hspace{0.2mm}\phi_f^2}\mp\frac{27
n^2\kappa^2}{2048\pi\hspace{0.2mm}G^5(V_0\phi_f^{n+1})^2}.
\end{eqnarray}
Thus the condition for ending inflation ($\epsilon(\phi_f) \simeq
1$) becomes,
\begin{eqnarray}\label{eqphi1}
\frac{n^2}{16\pi \hspace{0.2mm}G\hspace{0.2mm}\phi_f^2}\mp\frac{27
n^2\kappa^2}{2048\pi\hspace{0.2mm}G^5(V_0\phi_f^{n+1})^2}-1=0.
\end{eqnarray}
This equation has analytical solutions for $n=1$ and $n=2$. The
case $n=1$ has already been studied in \cite{Lamb} within the
Kaniadakis framework. In this work, we focus on the $n=2$ case and
compute both the Kaniadakis case and its dual. Therefore, all
subsequent calculations are performed for
 $n=2$. Solving  Eq. (\ref{eqphi1}) for  $\phi_f$, we obtain
\begin{eqnarray}\label{phic33}
\phi_f\simeq\frac{1}{2\sqrt{\pi
G}}\mp\frac{27\kappa^2}{32V_0^2}\sqrt{\frac{\pi^3}{G^5}}.
\end{eqnarray}
Based on Eq.~\eqref{efold1}, the value of $N$ can be obtained as
\begin{eqnarray}
    N\simeq 2 \pi G(\phi_c^2-\phi_f^2) \mp \hspace{0.2mm} \frac{9\pi \kappa^2}{64\hspace{0.2mm}G^3V_0^2(\phi_c^2-\phi_f^2) }\,.
\end{eqnarray}
Substituting $\phi_f$ from Eq.~\eqref{phic33}, the value of
$\phi_c$ is then determined as
\begin{eqnarray}
    \label{phic22}
    \phi_c\simeq\sqrt{\frac{1+2N}{4\pi\hspace{0.2mm}G}}\mp\frac{(6-N)\, 9\pi \kappa^2}{64 G^3 V_0^2}\sqrt{\frac{\pi G}{\left(1+2N\right)}}\,.
\end{eqnarray}
Using this relation, Eq. (\ref{eps}) can be rewritten in the
following form,
\begin{eqnarray}  \label{cc}
    \epsilon(\phi_c)&\simeq&\frac{1}{1+2N}\pm\frac{(11-2N) \,9\pi^2
        \kappa^2 N}{16\,G^2 V_0^2(1+2N)^3}.
\end{eqnarray}
Using the derivative from Eq.~\eqref{phii}, the parameter in
Eq.~\eqref{eta} for a power-law potential can be expressed as
\begin{eqnarray}  \label{eta2}
    \eta (\phi) \simeq  \frac{n(n-2)}{16\pi G \phi^2} \pm \frac{9n(3n+2)\kappa^2}{2048 \pi G^5 V_0^2 \phi^{2n+2}}.
\end{eqnarray}
In this equation, the standard leading-order term for the second
Hubble slow-roll parameter vanishes for $n=2$ (see Appendix A).
Thus, any non-zero dynamics for $\eta$ in this regime are sourced
by the Kaniadakis entropy corrections ($\kappa^2\neq0$). Therefore
Eq. \eqref{eta2} for $n=2$ can be rewritten as
\begin{equation}
    \eta (\phi_c)\simeq \pm \frac{9  \, \kappa^2 \, }{128 V_0^2 \pi G^5
    \phi_c^6}.
\end{equation}
Substituting $\phi_c$ from Eq.~\eqref{phic22}, leads to
\begin{eqnarray}
    \label{ccc}
    \eta(\phi_c)&\simeq&\pm\frac{9\pi^2\kappa^2}{2G^2 V_0^2\left(1+2N\right)^3}\,.
\end{eqnarray}
Inserting Eqs.~\eqref{cc} and \eqref{ccc} into Eqs.~\eqref{r} and
\begin{eqnarray}\label{rbis}
    r_k\simeq \frac{16}{1+2N} \pm \frac{9\pi^2 \kappa^2 (11-2N)N}{G^2
        V_0^2 (1+2N)^3}
\end{eqnarray}
\begin{eqnarray}\label{nn}
    n_{s_k} \simeq \frac{2N-3}{1+2N} \pm \frac{9\pi^2 \kappa^2 (2N^2 - 11N
        + 4)}{4G^2 V_0^2 (1+2N)^3}
\end{eqnarray}
The inflationary scenario proposed within the Kaniadakis-modified
Friedmann framework can be regarded as phenomenologically viable.
In Eq.~(\ref{nn}) for the scalar spectral index,
the $+$ and $-$ signs correspond to the Kaniadakis scenario and
its dual formation, respectively. By matching the theoretical
prediction for $n_s$ with its latest observational value from
BICEP and Planck ~\cite{ExpData}, \be \label{nsb} n_s = 0.965 \pm
0.004 \,\,\,\,\,\,\, (68\%\,\, \mathrm{CL})\,, \ee one can
estimate the parameter $\kappa$ within the framework of our
inflationary model. Assuming standard cosmological parameters,
such as $N \simeq 50$ e-folds and appropriate values for $G$ and
$V_0$, the analysis yields a consistent solution for the
Kaniadakis case, leading to an estimated value of the  parameter
$\kappa \simeq 4.1 \times 10^{-9}$. Here, $V_0 > 0$ is a constant
with dimension $[E]^{4-n}$, associated with the characteristic
energy scale of inflation, $E_{\mathrm{inf}} \simeq
\mathcal{O}(10^{15})\,\mathrm{GeV}$. In contrast,  the dual
formulation does not admit any physically admissible real solution
for $\kappa$ within the observationally allowed range of $n_s$,
implying that it is phenomenologically disfavored.
\begin{figure}[h]
\includegraphics[scale=0.95]{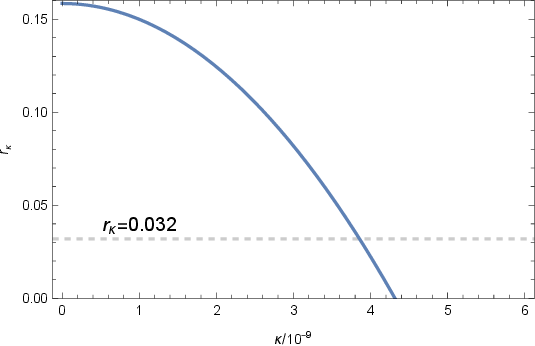}
\caption{The behaviour of $r_\kappa$ (Eq.~\ref{rbis}) versus the
(rescaled) $\kappa$ parameter for Kanidakis framework with
$V(\phi)=V_{0}\phi^{n}$ potential.}
    \label{fig1}
\end{figure}
    \begin{figure}[h]
\includegraphics[scale=0.95]{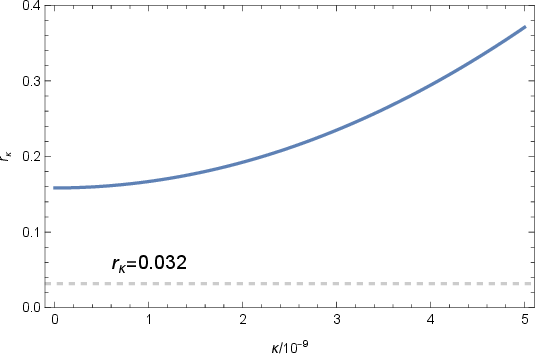}
\caption{The behaviour of $r_\kappa$ (Eq.~\ref{rbis}) versus the
(rescaled) $\kappa$ parameter for dual Kaniadakis framework with
$V(\phi)=V_{0}\phi^{n}$  potential.} \label{fig2}
\end{figure}
\\The tensor-to-scalar ratio $r_\kappa$, which we now proceed to
examine, provides an additional and independent observational test
of the model. According to the latest BICEP and Planck
observations~\cite{ConPlanck}, the tensor-to-scalar ratio is
constrained by \be r<0.032 \hspace{0.8cm} (95\%\,\,
\mathrm{CL})\,. \label{rb} \ee Figs.~\ref{fig1} and~\ref{fig2}
illustrate the behavior of $r_\kappa$ for the Kaniadakis and dual
formation, respectively. By comparing each branch with the
observational upper limit in Eq.~\eqref{rb}, we determine the
allowed range of the Kaniadakis parameter. Specifically, the
$\kappa$ parameter for the Kaniadakis branch is constrained to
values greater than $3.85 \times 10^{-9}$, consistent with the
constraint given in Eq.~\eqref{rb}.  In contrast, the dual branch
yields no physically valid solution.

In conclusion, the Kaniadakis-modified inflationary scenario is
consistent with current BICEP--Planck constraints, leading to a
small but physically admissible value of the parameter $\kappa
\gtrsim \mathcal{O}(10^{-9})$, whereas its dual formulation is
observationally disfavored due to the absence of real solutions
within the allowed ranges of $n_s$ and $r$. This motivates us to
further investigate the model in the next section within the
framework of the Mexican-hat potential
$V(\phi)=\lambda(\phi^{2}-\phi_{0}^{2})^{2}$.
\subsection{The case with potential $V(\phi)=\lambda (\phi^{2}-\phi_{0}^{2})^{2}$}
In this section, we first review the standard case of the Mexican
hat potential and its slow-roll inflationary behavior, and then
consider the corresponding scenario within the (dual) Kaniadakis
framework.
\subsubsection{ Standard slow-roll inflation with Mexican
hat potential} Using Eqs.~\eqref{fr2} and \eqref{phidot1}, we
obtain the following expression for the Mexican hat potential,
\begin{equation}
\label{phids} \dot\phi\simeq-2\sqrt{\frac{\lambda}{6\pi G}}\,\phi.
\end{equation}
 Upon substituting this result into Eq.~\eqref{eps22}, the following expression is derived
\begin{equation}
\label{bb} \epsilon(\phi)\simeq\frac{\phi^2}{\pi
\hspace{0.2mm}G\hspace{0.2mm}( \phi ^{2}-\phi_{0}^{2})^2}.
\end{equation}
As discussed in the previous sections, inflation ends when
$\epsilon(\phi_f) = 1 \,.$
This condition leads to
\begin{equation} \label{phifm}
\phi_f^2=\frac{\left( 2 \phi_0^2 + \frac{1}{\pi G} \right)\pm
\sqrt{\left( 2 \phi_0^2 + \frac{1}{\pi G} \right)^2-4
\phi_0^4}}{2}.
\end{equation}
An interesting feature arises for the potential $V(\phi) = \lambda
(\phi^2 - \phi_0^2)^2, $ where by setting the parameters $\phi_0 =
0$ and $\lambda=V_0$, this potential reduces to a power-law form
with $n = 4$. Therefore, by substituting $n=4$ into
Eq.(\ref{phif12}), we obtain $ \phi_f^2 \simeq \frac{1}{\pi G}. $
Comparing this expression with its limiting case, it follows that
the positive branch of the solution \eqref{phifm} should be
selected. Thus, we arrive at
\begin{equation}
\phi_f^2= \phi_0^2 + \frac{1}{2\pi G}+ \frac{1}{2\pi G} \sqrt{1+
4\pi G \phi_0^2 }.
\end{equation}
Assuming that $\phi_0^2 \ll M_{\rm Pl}^2=(8\pi G)^{-1}$, the
second term is much smaller than the first term, and therefore the
expression can be approximated as
 \begin{equation}
 \phi_f^2 \simeq 2\phi_0^2+ \frac{1}{\pi G}.
 \end{equation}
This equation can be rewritten in the following form
\begin{align}
\label{phif14} \phi_f\simeq\frac{1}{\sqrt{\pi G}}+ \phi_0^2
{\sqrt{\pi G}}\,.
\end{align}
Using Eq.~\eqref{efold1}, the number of e-folds $N$ can be
expressed as
\begin{align}
\label{efold} N = \int_{\phi_f}^{\phi_c} \frac{2\pi
G(\phi^2-\phi_0^2)}{\phi}\, d\phi.
\end{align}
Thus, we find
\begin{eqnarray}
    \label{ln}
    N\simeq  \pi G(\phi_c^2-\phi_f^2) -2\pi G
    \phi_0^2\ln\left(\frac{\phi_c}{\phi_f}\right).
\end{eqnarray}
First, we compute the term
$\ln\left(\frac{\phi_c}{\phi_f}\right)$. It is easy to show that
\begin{eqnarray}
\frac{\phi_c}{\phi_f}= \frac{\phi_c}{\frac{1}{\sqrt{\pi G}}+
\phi_0^2 {\sqrt{\pi G}}}= \sqrt{\pi G}\,\phi_c \, \left(1 - \pi G
\phi_0^{\,2}\right).
\end{eqnarray}
Consequently,
\begin{eqnarray}
    \ln\!\left(\frac{\phi_c}{\phi_f}\right)
    &=&
    \ln\!\left( \sqrt{\pi G\, \phi_c^{\,2}} \right)
    + \ln\!\left( 1 - \pi G \phi_0^{\,2} \right).
\end{eqnarray}
\noindent Assuming \(x \equiv \pi G \phi_0^2 \ll 1\), we employ
the Taylor expansion, $ \ln(1-x) = -x - \frac{x^2}{2} +
\mathcal{O}(x^3), $ to obtain the approximation
\begin{equation}
    \ln\!\left(\frac{\phi_c}{\phi_f}\right)
    \simeq \frac{1}{2}\ln(\pi G \phi_c^2) - \pi G \phi_0^2.
\end{equation}
Higher-order terms in the expansion are much smaller and can
therefore be neglected. Substituting this expression into
Eq.~(\ref{ln}), yields
\begin{eqnarray}
N & \simeq & \pi G (\phi_c^2-\phi_f^2)-\pi G \phi_0^2 \ln(\pi G
\phi_c^2),
\end{eqnarray}
where we have neglected the term $\mathcal{O}(\pi G \phi_0^2)^2$,
since we have assumed  $\pi G \phi_0^2 \ll 1$. Combining the
expression of $\phi_f$ from Eq.~\eqref{phif14}, we obtain
\begin{eqnarray}
    \label{num}
    N\simeq  \pi G\phi_c^2-1-\pi G
    \phi_0^2\left[2+\ln(\pi G\phi_c^2)\right].
\end{eqnarray}
To estimate the value of the inflaton at horizon crossing, we
start from the number of e-folds given in Eq. (\ref{num}). If the
coefficient $\pi G\phi_0^2$ is small enough that the logarithmic
term represents only a minor correction, the dominant contribution
comes from the first term, giving $\pi G\phi_c^2 \simeq N + 1$.

For the typical inflationary range $N \simeq 50\!-\!60$, this
yields $\pi G\phi_c^2 \simeq 51\!-\!61$. We then find $\ln(\pi
G\phi_c^2) \simeq \ln(N+1) \approx 4$, which is indeed of order
unity. The expression in brackets therefore becomes $2 + \ln(\pi
G\phi_c^2) \simeq 6$, confirming \textit{a posteriori} that the
logarithmic term provides only a perturbative correction, thus
validating our initial assumption. Under the above condition, we
obtain
\begin{align}
    \label{phic3}
    \phi_c\simeq\sqrt{\frac{N+1}{\pi\hspace{0.2mm}G}}+  3\phi_0^2 \sqrt{\frac{\pi\hspace{0.2mm}G}{N+1}}\,.
\end{align}
There is also another approach to deal with the problem and obtain
an explicit expression for $\phi_c$ in terms of $N$. In this case,
we first rewrite Eq. (\ref{num}) in the form
\begin{equation} \label{fx}
    f(x) = x - 1 - A(2 + \ln x)- N=0,
\end{equation}
where $ A \equiv \pi G \phi_0^2$ and $ x \equiv \pi G \phi_c^2$.
Now we solve Eq. (\ref{fx}) iteratively using a single
Newton--Raphson correction. Starting from the suitable initial
guess $x_0 = N + 1$, we find
\begin{align}
    f(x_0) &= -A[2 + \ln(N + 1)], \qquad
    f'(x_0) = 1 - {A}/(N + 1), \nonumber\\
    \delta x &= -\frac{f(x_0)}{f'(x_0)} =
    \frac{A[2 + \ln(N + 1)]}{1 - {A}/(N + 1)}, \\
    x &= x_0 + \delta x = (N + 1) +
    \frac{A[2 + \ln(N + 1)]}{1 - {A}/(N + 1)}. \nonumber
\end{align}
Therefore, the critical field becomes
\begin{align}
\phi_c= \sqrt{\frac{x}{\pi G}} =
            \left\lbrace \frac{N + 1}{\pi G} +
            \phi_0^2 \cdot
            \frac{2 + \ln(N + 1)}{1 -{\pi G \phi_0^2}/(N +
            1)}\right\rbrace^{1/2}
\end{align}
In the physically relevant limit $\pi G \phi_0^2 \ll N + 1$, the
denominator may be approximated as unity, leading to the
simplified form
\begin{equation}\label{phick}
    \phi_c \simeq
    \sqrt{
        \frac{N + 1}{\pi G}}+3\phi_0^2 \sqrt{
        \frac{\pi G}{N + 1}},
       \end{equation}
where we have used the previous result, $\ln(N+1)\simeq 4$, in the
last step. Therefore, the resulting expression for $\phi_c$
becomes consistent with the initial leading-order estimate given
in Eq.~(\ref{phic3}). This confirms that treating the logarithmic
term as a perturbative correction is self-consistent.

Using Eq.~(\ref{phic3}), we can compute the tensor-to-scalar ratio
and the scalar spectral index. Specifically, substituting
Eq.~(\ref{phic3}) into Eq.~(\ref{bb}), after some simplification,
we arrive at
\begin{eqnarray}
\epsilon(\phi_c) \approx \frac{1}{N+1} - \frac{4\pi G \phi_0^2}{(N+1)^2} .
\end{eqnarray}
\\By taking the derivative of Eq.~\eqref{phids}, Eq.~\eqref{eps223}, can be rewritten as follows
\begin{eqnarray}
{\eta(\phi_c) \simeq \frac{1}{2\pi G (\phi_c^2 - \phi_0^2)}}.
\end{eqnarray}
Substituting $\phi_c$ from Eq. \eqref{phick} into this expression,
yields
\begin{eqnarray}
\eta(\phi_c) \simeq \frac{1}{2(N+1)} - \frac{5\pi G  \phi_0^2 }{2(N+1)^2} .
\end{eqnarray}
In order to estimate the tensor-to-scalar ratio and the scalar
spectral index, we substitute these results into Eqs.~\eqref{r}
and \eqref{ns}. We find
\begin{equation}
    r \approx \frac{16}{N+1} - \frac{64 \pi G \phi_0^2}{(N+1)^2},
\end{equation}
\begin{equation}
    n_{s} \approx 1 - \frac{3}{N+1} + \frac{11\pi G \phi_0^2}{(N+1)^2}
\end{equation}
\subsubsection{Kaniadakis-modified slow-roll inflation Mexican
hat potential} Following the approach used to analyze the
Kaniadakis framework in Section 2.A, we now turn our attention to
the Mexican hat potential, examining the results of this
application.

Substituting Eq.~\eqref{Happ} into Eq.~\eqref{phidot1} and using
the potential, $V(\phi)=\lambda (\phi^{2}-\phi_{0}^{2})^{2}$,
yields the following expression
\begin{equation}
\dot\phi\simeq  \frac{-4 \lambda \phi (\phi^{2}-\phi_{0}^{2})}{3}
\left\lbrace \sqrt{\frac{8\pi G}{3}V} \pm \sqrt{\frac{27\pi}{2
G^{7} V^{3}}} \frac{\kappa^2}{64} \right\rbrace^{-1}.
\end{equation}
The resultant expression is as follows
\begin{equation}
\label{phidm} \dot\phi\simeq-2\sqrt{\frac{\lambda}{6\pi
G}}\phi\pm\frac{3 \phi k^2}{128 \,(\phi^{2}-\phi_{0}^{2})^{4}}
\sqrt{\frac{3 }{2\pi G^9 \lambda^3}}.
\end{equation}
Substituting this expression into Eq.~\eqref{eps}, we obtain
\begin{equation}
\begin{aligned}\label{ee}
\epsilon(\phi) = &\left\lbrace \frac{3}{2 \lambda(\phi^{2} -
\phi_{0}^{2})^2}
\mp \frac{27 \kappa^2}{128 G^4 \lambda^3 (\phi^2 - \phi_{0}^{2})^6} \right\rbrace \\
&\times \biggl( \frac{2\lambda \phi^2}{3\pi G} \mp \frac{3 \phi^2
\kappa^2}{64 \pi G^5 \lambda (\phi^{2}-\phi_{0}^{2})^{4}} \biggr).
\end{aligned}
\end{equation}
As previously mentioned, at the end of the inflation period, the
slow-roll parameter
 $\epsilon(\phi_f) $ reaches a value of 1. Therefore, Eq. \eqref{ee} above can be rewritten in the following form
\begin{equation}
\frac{\phi_f^2}{\pi \hspace{0.2mm}G\hspace{0.2mm}(
\phi_f^2-\phi_{0}^{2})^2}\,\mp\frac{27\, \phi_f^2 \,k^2 }{128\,
\pi\,  G^5 \lambda^2 \,(\phi_f^2-\phi_{0}^{2})^6}-1=0.
\end{equation}
To solve this equation, we first rewrite it in the following form
\begin{equation}
    \frac{aX}{(X - X_0)^2} \mp\frac{bX}{(X - X_0)^6} - 1 = 0,
\end{equation}
where $X=\phi_f^2$, $X_0=\phi_{0}^{2}$, $a=(\pi G)^{-1}$ and
$b=\frac{27\, \,k^2 }{128\, \pi\,  G^5 \lambda^2}$. Given the
complexity of the exact equation, we employ a perturbative
approach. A first-order Taylor expansion is applied to obtain an
approximate analytical solution. This yields the following result
for the square of the scalar field
\begin{equation}
X \simeq \frac{X_0^2}{a} \pm \frac{1}{a^2 X_0^2} \, b + \mathcal{O}(b^2).
\end{equation}
Substituting the original parameters into the solution, we derive
the explicit expression for the field in terms of the fundamental
constants of the model,
\begin{equation}
\phi_f^2 \simeq \pi G \phi_0^4 \pm \frac{27 \pi k^2}{128 G^3\lambda^2 \phi_0^4} + \mathcal{O}(k^4).
\end{equation}
Finally, we arrive at the following expression
\begin{equation}
\label{pi}
\phi_f \simeq \phi_0^2 \sqrt{\pi G} \pm \frac{27\,
k^2}{256  \lambda^2 \phi_0^6}\sqrt{\frac{ \pi}{G^7}}.
\end{equation}
Based on Eq. (\ref{efold1}), the number of e-folds can be
expressed as
\begin{align}
    \label{efold}
 N = \int_{\phi_f}^{\phi_c} \frac{8\pi G \lambda (\phi^2-\phi_0^2)^2 \pm \frac{9
 \pi\, k^2}{16 G^3  \lambda (\phi^2-\phi_0^2)^2}}{4 \lambda\phi (\phi^2-\phi_0^2)}\,
 d\phi.
\end{align}
Using Eq. (\ref{pi}), the approximate solution of this relation
becomes,
\begin{eqnarray} \label{eq:main1}
N &\simeq & \pi G \phi_c^2
        - (\pi G \phi_0^2)^2
        \left[1+ \frac{2}{\pi G \phi_0^2}\ln\!\left(\frac{\phi_c}{\phi_0^2 \sqrt{\pi G}} \right)\right]
        \nonumber\\
       &&\pm \frac{27 \pi^2 k^2 (1-\phi_0^2)}{128 G^2 \lambda^2
        \phi_0^6}.
\end{eqnarray}
To solve this equation, we assume that the second term inside the
parentheses including both its numerator and denominator is of the
same order. We will verify the validity of this assumption a
posteriori. Under this assumption, we obtain
    \begin{equation}
    N \simeq \pi G \phi_c^2
    - 2\pi^2 G^2 \phi_0^4
\pm \frac{27 \pi^2 k^2 (1-\phi_0^2)}{128 G^2 \lambda^2
    \phi_0^6}.
    \label{eq:main}
\end{equation}
Since $\pi^2 G^2 \phi_0^4\ll 1$, the second term can be neglected in above equation. Thus, we have
\begin{equation}\label{e4}
    \phi_c \simeq \sqrt{\frac{N}{\pi G}} \mp \frac{27  k^2 (1-\phi_0^2)}{256  \lambda^2 \phi_0^6}\sqrt{\frac{\pi^3}{N
    G^5}}.
\end{equation}
According to Eq. (\ref{ee}), $\epsilon(\phi_c)$ is obtained as
\begin{equation}
    \epsilon(\phi_c)\simeq  \frac{\phi_c^2}{\pi \hspace{0.2mm}G\hspace{0.2mm}(
        \phi_c^2-\phi_{0}^{2})^2}\,\mp\frac{27\, \phi_c^2 \,k^2 }{128\, \pi\,  G^5 \lambda^2
        \,(\phi_c^2-\phi_{0}^{2})^6}.
\end{equation}
Upon inserting Eq. (\ref{e4}), the expression becomes
\begin{equation}\label{e5}
    \epsilon(\phi_c)\simeq\frac{N}{\left(\pi G \phi_0^2 - N\right)^2}
\mp \frac{27 \pi^2 k^2  \left(\pi G \phi_0^2 + N\right)(1-\phi_0^2)}{128 G^2
\lambda^2 \phi_0^6 \left(\pi G \phi_0^2 - N\right)^3}.
\end{equation}
We now proceed to calculate $\eta$. As a first step, using Eq.
(\ref{phidm}), $ \ddot{\phi}$ is found to be
\begin{equation}
    \ddot{\phi}
    =
    \left[
    -2\sqrt{\frac{\lambda}{6\pi G}}
    \;\mp\;
    \frac{3k^2}{128}
    \sqrt{\frac{3}{2\pi G^9\lambda^3}}
    \frac{7\phi^2+\phi_0^2}{(\phi^2-\phi_0^2)^5}\right]\dot\phi.
\end{equation}
Substituting this result into Eq. (\ref{eta}) and performing the
necessary calculations, we obtain the following result
\begin{equation} \eta(\phi)\simeq
\frac{1}{2\pi G (\phi^{2}-\phi_{0}^{2})} \pm \frac{9 k^{2} (7\phi^2+\phi_0^2)}{256 \pi G^{5} \lambda^{2} (\phi^{2}-\phi_{0}^{2})^{6}}
\end{equation}
Thus, we arrive at
\begin{equation}
\label{e6}
\eta(\phi_c) \simeq
-\frac{1}{2\left(\pi G \phi_0^{2} - N\right)}
\pm\frac{27\pi^{2}k^{2}(1-\phi_0^{2})}{256\, G^{2}\lambda^{2}\phi_0^{6}\left(\pi G \phi_0^{2} - N\right)^{2}}.
\end{equation}
Inserting Eqs. (\ref{e5}) and (\ref{e6}) into Eqs. (\ref{r}) and (\ref{ns}) leads to the following results
\begin{equation} \label{rm}
    r_k = \frac{16N}{\left(\pi G \phi_0^2 - N\right)^2} \mp \frac{27 \pi^2 k^2 \left(\pi G \phi_0^2 + N\right)(1-\phi_0^2)}{8 G^2 \lambda^2 \phi_0^6 \left(\pi G \phi_0^2 - N\right)^3},
\end{equation}

\begin{equation} \label{nm}
    n_{s_k} = 1 - \frac{\pi G \phi_0^2 + 3N}{\left(\pi G \phi_0^2 - N\right)^2} \pm \frac{27\pi^2 k^2 \left(5\pi G \phi_0^2 + 3N\right)(1-\phi_0^2)}{128 G^2 \lambda^2 \phi_0^6 \left(\pi G \phi_0^2 - N\right)^3}.
\end{equation}
where as before, the upper and lower signs in the equations
correspond to the Kaniadakis and dual formulations, respectively.\\
 \begin{figure}
    \centering
    \includegraphics[scale=0.9]{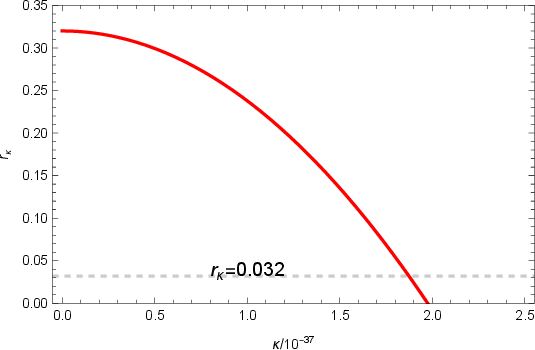}
    \caption{The behaviour of $r_\kappa$ (Eq.~\ref{rm}) versus the
        (rescaled) $\kappa$ parameter for the Kaniadakis framework with
        $V(\phi)=\lambda(\phi^{2}-\phi_{0}^{2})^{2}$ potential.}
    \label{fig3}
 \end{figure}

 \begin{figure}
    \centering
    \includegraphics[scale=0.9]{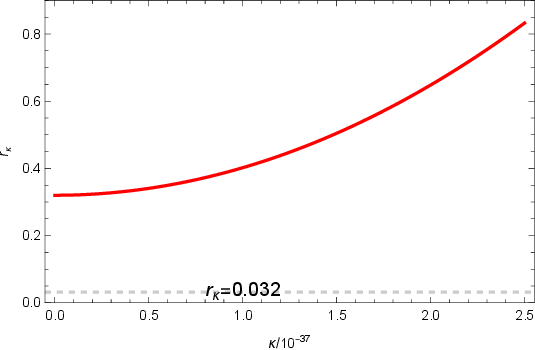}
    \caption{The behaviour of $r_\kappa$ (Eq.~\ref{rm}) versus the
        (rescaled) $\kappa$ parameter for the dual Kaniadakis framework with
        $V(\phi)=\lambda(\phi^{2}-\phi_{0}^{2})^{2}$ potential.}
    \label{fig4}
 \end{figure}
 Before proceeding with the analysis, it is worth noting that in
standard single-field inflation with symmetry-breaking potentials,
the quartic coupling is typically constrained to lie within the
range $\lambda \sim 10^{-13}$, as dictated by CMB normalization
\cite{Baumann}. Assuming $N \sim 50$ and $\phi_0^2 \ll M_{\rm
Pl}^2$, we now compare the predictions of Eqs.~(\ref{rm}) and
(\ref{nm}) against the latest observational constraints. By
matching Eq.~(\ref{rm}) with the tensor-to-scalar ratio from
Eq.~(\ref{rb}), we obtain an lower bound on the Kaniadakis
parameter, $\kappa \gtrsim 1.8 \times 10^{-37}$. Likewise,
comparing Eq.~(\ref{nm}) with the scalar spectral index  given in
Eq.~(\ref{nsb}) yields a consistent estimate of $\kappa \sim 2.12
\times 10^{-37}$. In both cases, only the Kaniadakis scenario
leads to physically acceptable solutions; the dual scenario fails
to satisfy the observational requirements. As a result, the lower
bound of the Kaniadakis parameter is naturally constrained to be
of the order of \(10^{-37}\), in agreement with current
observational bounds. The observational viability of the
Kaniadakis scenario is further supported by the analysis shown in
Figs. 3 and 4, which illustrate the behavior of the re-scaled
parameter $\kappa$ as a function of the tensor-to-scalar ratio $r$
for both Scenarios. These figures clearly demonstrate that only
the trajectory corresponding to the Kaniadakis scenario remains
within the observationally allowed region. In contrast, the dual
formation yields predictions that fall outside these bounds,
reinforcing the physical relevance and phenomenological
consistency of the Kaniadakis-modified inflationary scenario.

This result confirms that our earlier assumption namely, that the
numerator and denominator of the second term inside the
parentheses in Eq.~\eqref{eq:main1} are of the same order is
indeed valid. The consistency between this a posteriori
verification and the obtained solution reinforces the robustness
of our analysis.
 \section{Power spectrum and phenomenological consistency}\label{Power spectrum}
This section investigates the primordial power spectrum $P_\zeta$
for scalar perturbations under the Kaniadakis and dual Kaniadakis
entropy formalisms; these approaches are treated as alternative
phenomenological scenarios whose predictions are evaluated
independently against observational data to assess the model's
consistency. To quantify their impact on inflationary observables,
we focus specifically on the primordial scalar power spectrum. It
is therefore useful to first introduce a precise mathematical
definition of this quantity, which is characterized by the
two-point correlation function of the curvature perturbation
$\zeta$ as
\begin{equation}
    \langle \zeta^2(\vec{r}) \rangle = \int dk \,
    \frac{P_\zeta(k)}{k}.
\end{equation}
The energy scale \(k\) is related to the number of e-folds via the expression \cite{Arb}
\begin{equation}
    N(k) = \log\left(\frac{k_{\rm final}}{k}\right).
\end{equation}
The Harrison-Zeldovich (HZ) fit provides a simple description of
the CMB power spectrum ~\cite{HZ}
\begin{equation}\label{HZ}
    P_{\zeta}^{\rm HZ}(k) = A_s \left(\frac{k}{k_*}\right)^{n_s-1},
\end{equation}
where the characteristic amplitude $A_s$ is constrained to be $
\simeq 2.1* 10^{-9}$ at the pivot scale  $k_* =
0.05\,\mathrm{Mpc}^{-1}$. In the slow-roll approximation, the
primordial power spectrum can be written as $P_{\zeta}^{SR}\simeq
P_0 \log^2\left(\frac{k}{k_{\rm final}}\right)= P_0 N^2$, where
\(P_0\) is a constant. Furthermore, the power spectrum is directly
linked to the dynamics of the inflaton potential $V(\phi)$. In the
slow-roll approximation, this relationship is given by \cite{PV}
\begin{equation}
    \label{eq:PV_relation}
    P_\zeta(N) = \frac{V^2}{12\pi^2 M_{Pl}^4} \left( \frac{dV}{dN} \right)^{-1},
\end{equation}
which explicitly connects the observable spectrum to the evolution of the potential.

While the simple form of the primordial power spectrum, as given
by Eq.~\eqref{HZ}, offers a compelling fit to experimental data
using minimal parameters, it is not the optimal observational
match. According to \cite{deBlas}, the optimal observational match
is achieved by incorporating a running of the spectral index. This
capability allows the primordial power spectrum to deviate from
the simple power-law scaling at large cosmological scales and
undergo suppression at those scales. A significant refinement of
the primordial power spectrum Eq.~\eqref{HZ} can be obtained by
employing the multi-parameter broken-power-law (bpl) spectrum, as
defined in \cite{deBlas}
\begin{equation} \label{mpmod}
    P^{bpl}_\zeta = A_{low}
    \left(\frac{k}{k_*}\right)^{n_s^{bpl}-1+\delta} \quad (k \leq
    k_b).
\end{equation}
While the standard spectrum Eq.~\eqref{HZ} holds for scales $k >
k_b$, the bpl model applies for $k \leq k_b$. To ensure the
resulting spectrum is continuous at the scale ${k_b}$, the low-$k$
amplitude is explicitly set as  \(A_{\mathrm{low}} \equiv A_s
\left(k_b / k_*\right)^{-\delta}\). For completeness, we note that
the Planck collaboration also examined an alternative shape of
simple power-law multiplied by an exponential cut-off
\cite{deBlas} which is typically employed in scenarios where a
kinetic-energy-dominated phase follows the slow-roll stage
\cite{eg}. The best fit of the bpl model (Eq.~\eqref{mpmod}) to
the Planck TT+lowP data is achieved with the parameters
$n_s^{bpl}=0.9658$, $\log(k_b/\text{Mpc}^{-1})=-7.55$, and
$\delta=1.14$~\cite{deBlas}.
\\
To constrain the parameters of the generalized entropy model using
observational information, we establish a phenomenological correspondence
between the theoretically derived scalar spectral index and the bpl
parameterization. In the low-wavenumber regime, the bpl spectrum behaves as
\begin{equation}
    P_{\zeta }(k)\propto k^{n_s-1+\delta}.
\end{equation}
On the other hand, within the generalized entropy framework, the scalar
spectral index is modified according to\begin{equation}
    \label{eqform} n_{s_\kappa}^{(i)} = n_s^{(i)} + C_\kappa^{(i)},
    \qquad i=1,2
\end{equation}
The index $i=1,2$ serves to distinguish between the two scenarios
analyzed in Section III:  $i=1$ corresponds to Eq. \eqref{nn}, and
$i=2$ corresponds to Eq. \eqref{nm}. $n_s^{(i)}$ denotes the
standard spectral index in the $\kappa=0$ limit, and $
C_\kappa^{(i)}$ represents the (dual) Kaniadakis-induced
correction. Consequently, the theoretical
primordial spectrum takes the effective form
\begin{equation}
    P_{_\zeta }^{\rm th}(k)
    \propto
    k^{n_s^{(i)}-1+C_\kappa^{(i)}}.
\end{equation}
Comparing the power of \(k\) in the theoretical and phenomenological
expressions leads to the identification
\begin{equation}
    \delta\equiv C_\kappa^{(i)}.
\end{equation}
Therefore, the observationally preferred value of the bpl shift parameter
\(\delta\) can be used to constrain the magnitude of the Kaniadakis
deformation parameter. We can see that $C_\kappa^{(i)}$ plays a role
analogous to the $\delta$-shift in the broken power-law spectrum
of Eq.~\eqref{mpmod}.Now we focus on the case $i=1$, which corresponds to Eq.
\eqref{nn}. In this case, we have
\begin{equation}
    \delta = C_\kappa^{(1)}=\pm \frac{9\pi^2 \kappa^2 (2N^2 - 11N
        + 4)}{4G^2 V_0^2 (1+2N)^3}  .
\end{equation}
Given the observationally favored positive value for $\delta$ the
Kaniadakis entropy yields a physically admissible parameter
$\kappa$, whereas the dual formulation fails to satisfy this
specific phenomenological requirement. Using the values of $N$ and
$V_0$ fixed in Section III and $\delta = 1.14$ from~\cite{deBlas},
we obtain $\kappa \simeq 4.6 \times 10^{-9}$ within the
Kaniadakis framework.

Now we focus on the case $i=2$, which corresponds to Eq.
\eqref{nm}. In this case, we can write
\begin{equation}\label{delta}
    \delta = C_\kappa^{(2)}=\pm \frac{27\pi^2 \kappa^2 \left(5\pi G \phi_0^2 + 3N\right)(1-\phi_0^2)}{128 G^2 \lambda^2 \phi_0^6 \left(\pi G \phi_0^2 - N\right)^3}.
\end{equation}
Using the values for the parameters fixed in Section III and
comparing them with the $\delta$ value, the $\kappa$ parameter is
found to be approximately $\kappa \simeq 2.8 \times 10^{-37}$. In
this case, we use the Kaniadakis framework (+ sign in Eq.
\eqref{delta}) to ensure delta is positive, and thus the standard
Kaniadakis entropy lead to a physically reasonable result.
\section{Conclusions and discussions \label{Con}}
In this work, we have explored the implications of Kaniadakis
cosmology for the slow-roll inflationary epoch of the early
Universe by considering both the standard entropy formulation
($S_\kappa$) and its dual counterpart ($S^*_\kappa$). The analysis
was carried out within the framework of modified Friedmann
dynamics induced by Kaniadakis entropy, and the theoretical
predictions were confronted with current observational constraints
from the CMB. To investigate the inflationary behavior in detail,
we examined two representative inflationary potentials: the
power-law potential $V(\phi)=V_{0}\phi^{n}$ and the
symmetry-breaking Mexican-hat potential
$V(\phi)=\lambda(\phi^{2}-\phi_{0}^{2})^{2}$. Within the slow-roll
approximation, we derived the corresponding inflationary
parameters and evaluated the resulting predictions for the scalar
spectral index $n_s$ and the tensor-to-scalar ratio $r$. In
addition, we considered the phenomenological effects associated
with the running of the spectral index, $\delta$ , particularly in
the context of a broken power-law description of the primordial
spectrum.

Our results indicate that the standard Kaniadakis formulation
leads to viable inflationary scenarios compatible with present
observational bounds. In particular, the model admits physically
meaningful real solutions for the deformation parameter $\kappa$.
For the quadratic power-law potential ($V(\phi) \propto
\phi^{2}$), the observational constraints imply $\kappa   \gtrsim
10^{-9}$. On the other hand, the symmetry-breaking Mexican-hat
potential leads to a much stronger bound, restricting the
deformation parameter to extremely small values of order $\kappa
 \gtrsim 10^{-37}$. These results suggest that although Kaniadakis
corrections modify the cosmological dynamics, the deformation
parameter must remain very small in order to preserve
compatibility with the precise measurements of ($n_s \approx
0.965$) and the current limits on ($r$).

Overall, the positive correction term arising from the standard
Kaniadakis entropy produces a consistent modification of the
Friedmann equation, allowing the inflationary dynamics to remain
observationally viable. The analysis therefore supports the
possibility that generalized entropic frameworks such as
Kaniadakis statistics may provide a meaningful extension of
standard cosmological thermodynamics. Future investigations
including higher-order corrections, reheating dynamics, and more
precise cosmological observations may further clarify the role of
Kaniadakis entropy in the physics of the early Universe.

In contrast, the dual Kaniadakis formulation fails to provide
viable solutions. For both inflationary potentials considered in
this work, the mathematical structure of the slow-roll equations
does not yield any physically admissible real solutions for the
deformation parameter $\kappa$ within the observationally allowed
ranges of $n_s$ and $r$. To probe deeper into the phenomenological
robustness of the models, we analyzed their ability to reproduce
the running of the spectral index ($\delta \approx 1.14$) favored
by the broken power-law (bpl) fit to Planck data. The bpl model
suggests a deviation from a pure power law, characterized by the
relation $\delta = C_\kappa$ where $C_\kappa$ denotes the
Kaniadakis correction to the spectral index. Solving this
condition leads to resulting values of the deformation parameter
of order $\kappa \sim 10^{-9}$ for the power-law potential and
$\kappa \sim 10^{-37}$ for the symmetry-breaking Mexican-hat
potential.
\begin{table}[t]
    \centering
    \begin{tabular}{|c| c c|}
        \hline
        \, $|\kappa|$\,\, &\, Physical framework \, &\,  Ref.\, \\
        \hline
        \,$10^{-125}$\, &
        Baryon Acoustic Oscillations (BAO) & \cite{Hernandez-Almada:2021aiw} \\[1.5mm]
        \hline
        \,$10^{-125}$\, &
        CC+SNIa+BAO & \cite{Hernandez-Almada:2021aiw} \\[1.5mm]
        \hline
        \,$10^{-124}$\, & Cosmological constant (CC) & \cite{Hernandez-Almada:2021aiw} \\[1.5mm]
        \hline
        \,$10^{-124}$\, &
        Type Ia supernova (SNIa) & \cite{Hernandez-Almada:2021aiw} \\[1.5mm]
        \hline
        \,$10^{-123}$\, &
        Hubble data & \cite{CosmKan4} \\[1.5mm]
        \hline
        \,$10^{-123}$\, &
        Strong lensing systems & \cite{CosmKan4} \\[1.5mm]
        \hline
        \,$10^{-123}$\, &
        HII galaxies & \cite{CosmKan4} \\[1.5mm]
        \hline
        \,$10^{-83}$\, & Baryogenesis
        & \cite{Luciv} \\[1.5mm]
        \hline
    \end{tabular}
    \caption{Cosmological constraints on Kaniadakis parameter.}
    \label{Tab1}
\end{table}
It is worth noting that the constraints obtained in the present
inflationary analysis differ significantly from those derived from
late-time cosmological observations. Various studies have reported
extremely stringent bounds on the Kaniadakis deformation parameter
obtained from different cosmological probes. As noted in previous
studies, cosmological measurements across different epochs-
including Baryon Acoustic Oscillations (BAO), Type Ia supernovae
(SNIa), and baryogenesis- provide a wide range of constraints on
the Kaniadakis parameter $\kappa$, spanning from
$\mathcal{O}(10^{-125})$ up to $\mathcal{O}(10^{-83})$.
Table~\ref{Tab1} summarizes representative constraints on the
Kaniadakis parameter compiled from previous studies. The large
spread among these bounds, particularly between early-Universe
probes (inflation and baryogenesis) and late-time cosmological
observations (BAO and SNIa), may indicate that the Kaniadakis
parameter is not necessarily constant. Instead, it may exhibit a
running behavior, namely $\kappa=\kappa(t)$ or equivalently
$\kappa=\kappa(E)$. From the holographic perspective of Kaniadakis
entropy, entropy effectively counts the evolving degrees of
freedom of the Universe. Consequently, a decreasing function of
cosmic time-approaching $\kappa \to 0$ at present-could naturally
reconcile these apparently disparate constraints. In such a
scenario, $\kappa$ would attain larger values during the earliest
epochs of cosmic evolution (such as inflation or baryogenesis) and
become progressively suppressed at later times. This
interpretation is conceptually similar to discussions in Tsallis
thermodynamics and is consistent with expectations arising from
quantum field theory and renormalization-group flow.
\appendix
\section{}
In the standard slow-roll approximation, the Hubble slow-roll
parameters ($\epsilon_H$, $\eta_H$) can be related to the
potential slow-roll parameters ($\epsilon_V$, $\eta_V$) at leading
order as follows \cite{Baumann}
\begin{equation}
    \epsilon_H \simeq \epsilon_V, \qquad \eta_H \simeq \eta_V - \epsilon_V,
\end{equation}
where the potential slow-roll parameters are defined by
\begin{equation}
    \epsilon_V = \frac{M_{\rm pl}^2}{2} \left( \frac{V_{,\phi}}{V} \right)^2, \qquad \eta_V = M_{\rm pl}^2 \frac{V_{,\phi\phi}}{V}.
\end{equation}
Considering a general power-law potential $V(\phi) = V_0 \phi^n$, we can explicitly calculate these parameters
\begin{equation}
    \epsilon_V = \frac{M_{\rm pl}^2}{2} \left( \frac{n}{\phi} \right)^2 = \frac{n^2 M_{\rm pl}^2}{2\phi^2},
\end{equation}
\begin{equation}
    \eta_V = M_{\rm pl}^2 \frac{n(n-1)}{\phi^2}.
\end{equation}
Substituting these into the relation for $\eta$ yields
\begin{equation}
    \eta_H \simeq \eta_V - \epsilon_V = M_{\rm pl}^2 \left[ \frac{n(n-1)}{\phi^2} - \frac{n^2}{2\phi^2} \right] = \frac{n(n-2) M_{\rm pl}^2}{2\phi^2}.
\end{equation}
Using the reduced Planck mass relation $M_{\rm pl}^2 = \frac{1}{8\pi G}$, this expression exactly reproduces the standard term in our Eq.~(\ref{eta2})
\begin{equation}
\eta_H \simeq \frac{n(n-2)}{16\pi G \phi^2}.
\end{equation}
This clearly shows that for the special case of a massive scalar
field potential where $n=2$ (i.e., $V(\phi) \propto \phi^2$), the
leading standard term completely vanishes ($\eta_{H}^{\rm
standard} = 0$). Consequently, in our modified framework, the
non-zero dynamics of $\eta_H$ in this regime are solely governed
by the entropy corrections (i.e., terms proportional to
$\kappa^2$).

\end{document}